\documentclass[11pt,a4paper,fleqn]{article}
\catcode`\@=11
\@addtoreset{equation}{section}
\def\theequation{\arabic{section}.\arabic{equation}}

\def\appendix{\renewcommand{\thesection}{\Alph{section}}\setcounter{section}{0}
              \renewcommand{\theequation}
            {\mbox{\Alph{section}.\arabic{equation}}}\setcounter{equation}{0}}

\def\maketitle{\thispagestyle{empty}\setcounter{page}0\newpage
                \renewcommand{\thefootnote}{\arabic{footnote}}
                  \setcounter{footnote}0}
\renewcommand{\thanks}[1]{\renewcommand{\thefootnote}{\fnsymbol{footnote}}
               \footnote{#1}\renewcommand{\thefootnote}{\arabic{footnote}}}

\renewcommand{\title}[1]{\begin{center}\Large\bf #1\end{center}\rm\par\bigskip}
\renewcommand{\author}[1]{\begin{center}\Large #1\end{center}}
\newcommand{\address}[1]{\begin{center}\large #1\end{center}}
\newcommand{\pacs}[1]{\smallskip\noindent{\sl PACS numbers:
                       \hspace{0.3cm}#1}\par\bigskip\rm}
\def\babs{\hrule\par\begin{description}\item{Abstract: }\it}
\def\eabs{\par\end{description}\hrule\par\medskip\rm}
\renewcommand{\date}[1]{\par\bigskip\par\sl\hfill #1\par\medskip\par\rm}
\newcommand{\ack}[1]{\par\section*{Acknowledgements} #1}

\def\dinfn{Dipartimento di Fisica, Universit\`a di Trento\\
                           and Istituto Nazionale di Fisica Nucleare,\\
                                   Gruppo Collegato di Trento, Italia \medskip}

\def\csic{Consejo Superior de Investigaciones Cient\'{\i}ficas \\
Instituto de Ciencias del Espacio (ICE/CSIC)\\
 Campus UAB, Facultat de Ci\`encies, Torre C5-Parell-2a planta, \\
08193 Bellaterra (Barcelona), Spain \medskip}

\def\ieec{Institut d'Estudis Espacials de Catalunya (IEEC), \\
Edifici Nexus, Gran Capit\`a 2-4, 08034 Barcelona, Spain \medskip}

\def\guido{Guido Cognola\thanks{e-mail: \sl cognola@science.unitn.it\rm}}
\def\sergio{Sergio Zerbini\thanks{e-mail: \sl zerbini@science.unitn.it\rm}}

\def\emilio{Emilio Elizalde\thanks{e-mail: \sl elizalde@ieec.uab.es\rm}}

\def\hs{\qquad}               
\def\nn{\nonumber}            
\def\beq{\begin{eqnarray}}    
\def\eeq{\end{eqnarray}}      
\def\ap{\left.}               
\def\at{\left(}               
\def\aq{\left[}               
\def\ag{\left\{}              
\def\cp{\right.}              
\def\ct{\right)}              
\def\cq{\right]}              
\def\cg{\right\}}             
\def\R{{\hbox{{\rm I}\kern-.2em\hbox{\rm R}}}}   
\def\H{{\hbox{{\rm I}\kern-.2em\hbox{\rm H}}}}   
\def\N{{\hbox{{\rm I}\kern-.2em\hbox{\rm N}}}}   
\def\C{{\ \hbox{{\rm I}\kern-.6em\hbox{\bf C}}}} 
\def\Z{{\hbox{{\rm Z}\kern-.4em\hbox{\rm Z}}}}   
\def\ii{\infty}                                  
\def\Tr{\mathop{\rm Tr}\nolimits}                  
\def\Res{\mathop{\rm Res}\nolimits}                
\renewcommand{\Re}{\mathop{\rm Re}\nolimits}       
\def\dir{/\kern-.7em D\,}                            
\def\lap{\Delta\,}                                 
\def\al{\alpha}
\def\be{\beta}
\def\ga{\gamma}

\def\ze{\zeta}

\def\la{\lambda}

\def\om{\omega}

\def\Ga{\Gamma}

\def\Om{\Omega}

\begin{document}


\title{Heat-kernel expansion on non compact domains and a generalised
zeta-function regularisation procedure}

\author{\guido$^{,a}$, \emilio$^{,b,c}$ and \sergio$^{,a}$}
\address{\small $^a$\dinfn \\ $^b$\csic \\ $^c$  \ieec}

\babs Heat-kernel expansion and zeta function regularisation are
discussed for Laplace type operators with discrete spectrum in non
compact domains. Since a general theory is lacking, the heat-kernel
expansion is investigated by means of several examples.
It is  pointed out that for a class of exponential (analytic)
interactions,  generically the non-compactness of the domain 
gives rise to
logarithmic terms in the heat-kernel expansion. Then, a meromorphic
continuation of the associated zeta function is investigated. A
simple model is considered, for which the analytic continuation of the
zeta function is not regular at the origin, displaying a pole of
higher order. For a physically meaningful evaluation of the
related functional determinant, a generalised zeta function
regularisation procedure is proposed. \eabs

 \noindent Keywords: Heat-kernel
expansion, Zeta function-regularisation, functional determinants,
confining potentials.

\section{Introduction}
\label{Form}

Within the so-called one-loop approximation in quantum field theory,
the Euclidean one-loop effective action can be expressed in terms
of the sum of the classical action and a contribution depending on a
functional determinant of an elliptic differential operator, the so
called fluctuation operator. The ultraviolet one-loop divergences
which are present need to be regularised by means of a suitable
technique (for recent  reviews, see \cite{eliz94b}--\cite{byts03b}).

In general, one works in Euclidean spacetime and
deals with a self-adjoint, non-negative, second-order differential
operator of the form \beq L=-\lap+V\,, \eeq where $\lap$ is the
Laplace-Beltrami operator and $V$ a potential depending on the
classical  background solution and containing, in general, a mass
term. It is well known that the one-loop effective action 
$W\equiv W[\Phi]$, is related to the functional 
determinant of the field operator $L$ by \beq W=-\ln
Z=S_c+\frac{1}{2}\ln\det\frac{L}{\mu^2}\,, \label{I} \eeq 
$S_c$ being the classical action and $\mu^2$ a renormalisation
 parameter, which appears for dimensional reasons.

The one-loop divergences may be  dealt with
by using a variant of the zeta-function regularisation method
\cite{z1}-\cite{dowk76-13-3224}. One namely introduces the regularisation 
parameter $\varepsilon$ and considers
 \beq
W(\varepsilon)=S-\frac{1}{2}\int_0^\ii dt\
\frac{t^{\varepsilon-1}}{\Gamma(1+\varepsilon)} \Tr
e^{-tL/\mu^2}=S-\frac{1}{2\varepsilon}\ze(\varepsilon|L/\mu^2)\,,
\label{bb} \eeq where, as usual, for the elliptic operator $L$ the
zeta function is defined by means of the Mellin-like transform 
\beq
\zeta(s|L)=\frac{1}{\Ga(s)}\int_0^\ii dt\ t^{s-1} \Tr e^{-tL}\,,
\hs\hs \ze(s|L/\mu^2)=\mu^{2s}\ze(s|L)\:. 
\label{mt}\eeq 
Here the  heat trace $\Tr e^{-tL}$ plays a preeminent
role. Recall that, for a second-order elliptic non-negative operator
$L$ in a {\it compact} $d-$dimensional manifold without boundary, one has
the small-$t$ asymptotic expansion 
\beq \Tr
e^{-tL}\simeq\sum_{j=0}^\ii A_j(L) \, t^{j-d/2}\:, \label{tas00} \eeq
where $A_j(L)$ are the Seeley-DeWitt coefficients
\cite{dewi65b,seel67-10-172}. As a result,  for a
second-order differential operator in $d-$dimensions, 
the integral (\ref{mt}) is convergent in the domain 
$\Re s>\frac{d}{2}$.

In the compact case, $\zeta(s|L)$ is regular at the origin 
and one has the well known result $\zeta(0|L)=A_{d/2}(L)$.   
The latter quantity is  computable (see for example the recent reviews
\cite{kirsten00,vass03}) and depends  on the potential and on
 geometric invariants. In particular, for odd dimensional 
manifolds without boundaries, $\zeta(0|L)=0$. 
Performing a Taylor expansion of the zeta function we obtain 
\beq W(\varepsilon) =
S-\frac{1}{2\varepsilon}\zeta(0|L)+\frac{\zeta(0|L)}{2}\ln \mu^2+
\frac{\ze'(0|L)}{2}+ O(\varepsilon)\,. \label{bbb1} \eeq
Thus, the one-loop divergences as well as finite
contributions to the one-loop effective action are expressed 
in terms of the zeta function and its derivative evaluated at 
the origin.

In this paper, we would like to discuss a more general situation
where {\it logarithmic terms} in the heat-trace asymptotics are present.
First, we recall a well known but crucial fact concerning  the local 
heat-kernel expansion  associated with a Laplace type operator in  
R$^d$ of the kind 
\beq H=-\lap +V( x)\,. \eeq 
If the potential is real
and non-negative, with an additional, rather mild hypothesis, the
operator $H$ is essentially self-adjoint in $C_0^\ii($R$^d)$. We will
be interested in confining potentials which, with the additional
hypothesis of being also smooth functions,
 give rise to a discrete spectrum.
 As has been shown in Refs.~\cite{parker1,parker2}, the local 
heat-kernel expansion can be
partially summed over and rewritten under the form 
\beq
K_t(x,x)= \frac{1}{(4\pi t)^{d/2}}e^{-tV(x)}\sum_{n=0}^\ii
b_n(x)\, t^n\,, 
\label{pt} 
\eeq 
where the new coefficients $b_n(x)$ can
easily be  computed and depend only on the derivatives of the
potential $V(x)$. The first few read
\beq
&& b_0(x)=1\,,\hs\hs\hs b_1(x)=0\,, \nn
\\ 
&& b_2(x)=-\frac{1}{6}\lap V\,,\hs\hs
b_3(x)=-\frac{\lap^2V}{60}+\frac{\nabla_kV\nabla_kV}{12}\,,\label{G0}
\\
&& b_4(x)=-\frac{\lap^3 V}{840}+\frac{(\lap V)^2}{72}
+\frac{\nabla_i\nabla_jV\nabla_i\nabla_jV}{90}
+\frac{\nabla_kV\nabla_k\lap V}{30}
\,.
\label{G}
\eeq
We will make use of  such a re-summation for obtaining the
heat-kernel trace asymptotics in the next section.

If one is dealing with smooth compact manifolds, the passage to the 
heat-kernel trace is accomplished by integrating term by term over 
the coordinates, and no 
logarithmic contribution in the heat-trace expansion appears. 
However, in the case of non-smooth manifolds one may get
logarithmic terms in the heat-kernel trace, for
example when one considers the Laplace operator on higher-dimensional
cones \cite{bord96, cogn97}, but also in 4-dimensional
spacetimes with a 3-dimensional, non-compact, hyperbolic spatial
section of finite volume \cite{byts97}, and in the case of general
pseudo-differential operators \cite{grubb}. More recently, the
presence of logarithmic terms in self-interacting scalar field
theory defined on manifolds with non-commutative coordinates has
also  been pointed out \cite{byts01}-\cite{byts02}. This goes
together with a non-typical behaviour of the corresponding zeta
function: possibly a simple pole at the origin and higher-order
poles at other places.

Here we would like to investigate the case of Laplace-type
self-adjoint operators defined on {\it non-compact} manifolds. To
our knowledge, for the general case of a confining potential and
discrete spectrum, a systematic theory has  yet
to be formulated. A pioneering investigation along this line can be
found in \cite{nash}. Other studies involve one dimensional 
problems on the real half-line \cite{voros} and the Barnes zeta
functions \cite{dowker}.
With regard to the presence of logarithmic terms in heat-trace 
asymptotics, one should note that they have been considered 
in the abstract context of regularised products in many places. 

Recall that, under certain conditions, the regularised 
product associated with an infinite sequence of  non-zero 
complex numbers $\ag \lambda_n \cg$ has a related Dirichlet  series 
$\sum_n \lambda_n^{-s}$  (the zeta-function). In this paper, 
we are only interested in the case when the $\lambda_n$ are 
eigenvalues of
{\it non}-negative differential operators and the zeta function 
converges absolutely for 
$\Re s$ sufficiently  large. When this zeta-function is holomorphic 
at the origin, the regularised product is defined as 
$\exp{[ -\zeta'(0)]}$.  A general theory is presented in 
\cite{Illies,jorgenson} and other relevant 
papers are \cite{manin}-\cite{simon} and references quoted therein.
It is worth mentioning, that the case of non-compact domains
but with scattering potentials, namely the ones for
which  a continuous spectrum exists, is well understood and the
 $S-$matrix or the phase shift function then enter the game
(see, for instance, \cite{muller}).
In this context, delta-like potentials have also been
considered (see for example \cite{solo}-\cite{nail}, and references
therein). If the potential is singular, for
instance, proportional to $1/x^2$, the presence of logarithmic
terms in the local heat-kernel expansion is also possible, their
coefficients becoming distributions \cite{callias} (see also the
 recent paper \cite{kirsten} and
references therein). Here, we will not deal with situations
of this kind. Moreover, for the sake of simplicity, we will limit
ourselves to the R$^d$ flat case.

The content of the paper is as follows. In Sect.~2, local
heat-kernel asymptotics are reviewed for the simple case we are going
to deal with. In Sect.~3, the heat-trace asymptotics are
investigated and the possibility for the presence of logarithmic
terms is pointed out explicitly. The consequences of such unusual
terms are discussed in Sect.~4 in some detail. Finally, in Sect.~5,
a simple model of confinement is proposed and a generalisation of the
zeta-function regularisation method is constructed to deal with
 this case. The paper ends with some conclusions and an Appendix.

\section{Heat-kernel trace asymptotics in non-compact domains}

In this section we will show that, for suitable classes of
potentials in non-compact domains, logarithmic terms can actually be
present in the heat-trace expansion. 
Under the usual hypothesis concerning the potential $V(x)$ 
---namely $V(x)$ smooth enough, non-negative and going to infinity 
for $|x|\to\ii$--- the heat trace can be shown to exist and
the heat kernel asymptotics is given by
\beq \Tr e^{-t H}=\int_{R^d} dx\,
K_t(x,x)= \frac{1}{(4\pi t)^{d/2}} \int_{R^d} dx\ e^{-tV(x)}\left[
1+t^2 b_2(x)+t^3b_3(x)+ \cdots \right]
\label{ptt} \eeq 
In particular we shall focus our attention on 
the class of spherical potentials, $V(x)=V(r)$, $r\in[0,\ii)$,
and thus
\beq \Tr
e^{-t H}= \frac{\Omega_d}{(4\pi t)^{d/2}} 
\int_0^\ii  dr\,\ r^{d-1}
e^{-tV(r)}\left[ 1+O(t^2)\right] \,, 
\label{pttr} \eeq where
$\Omega_d=2\pi^{d/2}/\Gamma(d/2)$. The latter expression will be
our starting point for further discussions.

As a first family of potentials, let us consider $V(r)$ to be a
positive polynomial of degree $Q$.
In this case, we will show that
logarithmic terms are absent, but the leading term goes as
$O(t^{-d/2-d/Q})$, in contrast to the leading behaviour $O(t^{-d/2})$
associated with the compact case.
To prove this, since we are interested in the short-$t$
leading term, it is sufficient to consider the leading
term of the potential, namely $V(r)=r^Q+\cdots$. 
Thus, the leading term in the heat trace reads
\beq 
\Tr e^{-tH}\simeq\frac{\Omega_d}{(4\pi t)^{d/2}}\,
\int_0^\ii  dr\,r^{d-1}
e^{-t\,r^Q}=
\frac{2^{1-d}\Gamma(d/Q)}{Q\,\Gamma(d/2)\, t^{d/2+d/Q}}
\,. \label{ptho1} \eeq
One can check that this result holds true for the case
$Q=2$,  for which the heat trace is well known, since it
corresponds to
the partition function of a harmonic oscillator in $d-$dimensions. In
fact one has eigenvalues $2n+1$ for each dimension and then
\beq \Tr
e^{-t H}=\at \sum_0^\ii e^{-t(2n+1)}\ct^d= \frac{1}{(2\sinh
t)^d}\simeq \frac{1}{(2 t)^d}+\cdots \,, \label{ho} \eeq 
in agreement with Eq.~(\ref{ptho1})
(note that here $m=1/2$ and $\om=2$). 
Our results also agree with those for the 1-dimensional
case investigated in \cite{voros}.

The situation drastically changes if one considers exponential
confining potentials which,  for large $r$, go asymptotically as
$V(r)\simeq e^{r^Q}$. We will show that in these cases logarithmic
terms are present.
In fact, we have
\beq \Tr e^{-tH}&\simeq& 
\frac{\Omega_d}{(4\pi)^{d/2}t^{d/2}}
\int_0^\ii dr\,r^{d-1}e^{-te^{r^Q}}
\nn\\
&=&\frac{\Omega_d}{(4\pi)^{d/2}\, t^{d/2}Q}
\int_1^\ii dy \ y^{-1} e^{-ty} \at\ln y\ct^{d/Q-1}\,.
\label{ex} \eeq
For simplicity, let us now assume $\frac{d}{Q}$ to be an integer. 
In such  case
we can use (\ref{XXX}) and (\ref{bbb2}) of the Appendix, thus
obtaining the leading term in the form
\beq 
\Tr e^{-t H}\simeq \frac{(-1)^{d/Q}}
{2^dQ\Gamma(d/2+1)\, t^{d/2}}\,(\ln t)^{d/Q} \,. 
\label{ex10} 
\eeq 
It has to be noted that with respect to the compact case,
for such class of potentials on non-compact manifolds,
the leading term in the trace is modified by the presence of 
the logarithmic factor $(\ln t)^{d/Q}$. We also note that,
for $Q=1$, (\ref{ex10}) yields the same result obtained 
by Nash in \cite{nash} using a different method.
Let us emphasise that those comparisons are essential both for 
consistency reasons and in view of its application to real 
situations in physics.

Equations (\ref{ptho1}) and (\ref{ex}) give only the leading term
in the trace of the heat kernel, but in principle it is possible to 
go on in the expansion by integrating other terms of the local
asymptotics. However it should be stressed that more terms in the 
local expansion can give contributions of the same order to the trace
asymptotics. This can be easily seen by considering,
for instance, the 1-dimensional harmonic oscillator
described by the Hamiltonian
\beq 
H=-\frac{d^2}{dx^2}+\frac{\om^2x^2}4\,,\hs\hs \hbar=1\,,\hs 
m=\frac12\,.
\label{}\eeq 
For this model, one  has
\beq 
\Tr e^{-tH}=\frac{1}{2\sinh(\om t/2)}
=\frac{1}{\om t}-\frac{\om t}{24}+O(t^3)
\label{OA}\eeq
\beq 
K_t(x,x)&=&\sqrt{\frac{\om}{4\pi\sinh\om t}}\,
  e^{-\om x^2(\cosh\om t-1)/2\sinh\om t}\nn\\
&=&\frac{e^{-tV}}{\sqrt{4\pi t}}\,\at1+b_2t^2+b_3t^3+\cdots\ct\,,
\nn\\&&\hs\hs
b_2=-\frac{\om^2}{12}\,,\hs\hs
b_3=\frac{\om^4x^2}{48}\,.
\label{}\eeq
In order to get the expansion (\ref{OA}) up 
to order $t$, one needs to integrate the local expansion up to order
$t^{5/2}$. This means that both $b_2$ and $b_3$ give a contribution
of order $t$ in the trace asymptotics. To obtain the subsequent 
term $t^3$, one has to consider all $b_n$ coefficients up to $b_6$.

\section{Meromorphic extension of the zeta-function}

With respect to the compact case, the meromorphic 
structure of the zeta function associated with the operator $H$
is generically quite complicated and it is strictly related to
the form of the potential. In order to show this, we first consider
the polynomial case $V(r)=r^Q$ and assume 
$Q=2P$ to be an even number. Under such assumption
all $b_n$ coefficients are polynomials in $r$ and 
the heat-trace asymptotics are of the form
\beq
\hspace*{-4mm} \Tr e^{-t H}=\sum_n\,C_n \, t^{\al_n-(d/2+d/Q)}\,,\hs
C_0=\frac{\Gamma(d/Q)}{2^{d-1}Q\Gamma(d/2)}\,,\hs
\al_0=1<\al_1<\al_2<\cdots
\label{ptho12}\eeq
where the $C_n$ are numerical coefficients obtained by integrating
the local expansion, and the $\al_n$ are rational numbers.
Making use of (\ref{mt}) and splitting the integration
over $t$ into $(0,1]$ and $[1,\ii)$, we get
\beq
\zeta(s|H)=\frac{1}{\Gamma(s)}
\sum_n \frac{C_n}{s+\al_n-(\frac{d}{2}+\frac{d}{Q})}+
\frac{J(s)}{\Gamma(s)}\,,
\label{}\eeq
where $J(s)$ is an analytic function.
It follows that for such class of potentials 
the zeta function admits only simple poles 
---as it happens in the compact case---
but whose location strictly depends on the form of the potential,
since the $\al_n$ are not universal powers.
Moreover, we see that $\ze(0|H)$ is not vanishing if and only if
$\al_n=d/2+d/Q$ for some $n$ and the corresponding coefficient $C_n$ 
is different from zero (note that if this coefficient $C_n=0$, then 
the corresponding term is absent from the sum, for any $s$). 

The situation becomes  more complicated for the class of
exponential potentials we have considered in the previous section.
In fact, in such case one obtains in general an asymptotic
expansion with
terms of the kind $t^{\al}(\ln t)^\be$, $\al$ and $\be$
being rational numbers which depend on the potential, and this means
that the meromorphic extension of the zeta function will have
poles or branch points of order $\be$ at $s=-\al$ 
(see (\ref{a1}) in the Appendix). 

In order to compute the non-holomorphic structure of the zeta function for 
this 
class of potentials it is convenient to proceed as follows.
We use the general expression (\ref{pt}) in (\ref{mt}) and thus,
for $\Re s$ sufficiently large and $V(x)=V(r)>0$, we can write
\beq 
\zeta(s|H)&\sim&\frac{1}{(4\pi)^{d/2}\Ga(s)}\,\sum_n
\int_{0}^{\ii}\,dt\,t^{s+n-d/2-1}\int_{R^d}\,dx\,b_n(x)e^{-tV(x)}
\nn\\
&=&\sum_n\,\frac{\Ga(s+n-d/2)}{(4\pi)^{d/2}\Ga(s)}\,
\int_{R^d}\,dx\,b_n(x)[V(x)]^{-(s+n-d/2)}\,,
\label{zetaY}\eeq
which is well defined for even $Q$, since in such case all 
coefficients $b_n(x)$ are regular everywhere.
Since $V(r)$ is exponential like and spherically symmetric, we may 
assume that
\beq 
b_n=\sum_{pq}\,C^n_{pq}r^{p}V^q\,,
  \hs0\leq p\leq2(n-1)(Q-1)\,,\hs 1\leq q<n\,,\hs n\geq2\,.
\label{bn}\eeq
Now, the integration can be performed and we obtain for the 
non-holomorphic part
\beq
\ze(s|H)&\sim&\frac{\Om_d}{(4\pi)^{d/2}\Ga(s)}\,\aq
\Ga(s-d/2)\,\int_0^\ii\,dr\,r^{d-1}e^{-(s-d/2)r^Q}
                \phantom{\sum_{n\geq2;p;q}}
\cp\nn\\&&\ap\hs
+\sum_{n\geq2;pq}\,C^n_{pq}\Ga(s+n-d/2)
\int_0^\ii\,dr\,r^{d+p-1}e^{-(s+n-q-d/2)r^Q}
\cq
\nn\\
&=&\frac{\Om_d}{(4\pi)^{d/2}Q\Ga(s)}\,\aq
\frac{\Ga(s-d/2)\Ga(d/Q)}{(s-d/2)^{d/Q}}
               \phantom{\sum_{n\geq2;p;q}}
\cp\nn\\&&\ap\hs
+\sum_{n\geq2;pq}\,C^n_{pq}
\frac{\Ga(s+n-d/2)\Ga((d+p)/Q)}{(s+n-q-d/2)^{(d+p)/Q}}
\cq\,.
 \label{ZEX}\eeq
As a consequence, it follows that generically  
the zeta function may have poles and branch points
of any order. It should also be noted that in the case of even 
dimension it is not holomorphic at the origin. 

For example, in the simplest case $d=Q=2$, $V(r)=e^{r^2}$,  
by straightforward dimensional analysis one 
can see that only the term proportional to $C^2_{21}$ 
contributes to the singularity at the origin and we get 
(see the next section)
\beq 
\ze(s|H)=\frac{C^2_{21}}{4s}+\cdots\,,\hs
\hs C^2_{21}=-\frac{2}{3}\,.
\label{}\eeq

We conclude this section by studying 
the asymptotics of the spectral density
associated with the operator $H$. We can define the spectral density
via the spectral representation of the heat trace, namely 
\beq 
\Tr e^{-t H}=\int_0^\ii  e^{-t \lambda}\,dN(\lambda)
= \int_0^\ii\,d\lambda e^{-t \lambda} \rho(\lambda)\,. 
\eeq 
For the polynomial interaction, 
the Tauberian theorems (see the Appendix) 
and the short-$t$ leading terms of the heat-trace
expansion give
\beq 
N(\lambda)\simeq
\lambda^{\frac{dQ+2d}{2Q}}\,, \hs \rho(\lambda)\simeq
\lambda^{\frac{dQ+2d}{2Q}-1}\,,\hs \lambda \rightarrow \ii\,,
\label{ap} \eeq 
while for the exponential interaction, with $d/Q$ an integer, 
\beq 
N(\lambda)\simeq \lambda^{\frac{d}{Q}} \,(\ln
\lambda)^{d/Q}\,,\hs \rho(\lambda)\simeq \lambda^{\frac{d}{Q}-1}
\,(\ln \lambda)^{\frac{d}{Q}} \,, \hs  \lambda \rightarrow \ii\,.
\label{ae} \eeq 
In particular, when $Q=d$, one has 
\beq N(\lambda)\simeq \lambda
\,(\ln \lambda)\,,\hs \rho(\lambda)\simeq  \, \ln \lambda \,, \hs
\lambda \rightarrow \ii\,. \label{aeprimi} 
\eeq 
In  this last case the distribution of the eigenvalues 
of the operator $H$ resembles the asymptotic behaviour 
which one meets in number theory,
namely the asymptotic distribution of the non-trivial zeroes of the
Riemann zeta function \cite{nash}. 
With regard to this important issue we refer the reader 
to the literature, mentioning  the relevance of the method based on 
Cramer's V-function \cite{cramer} 
and references therein. Other related papers are 
\cite{rosu}-\cite{sierra}.

\section{A simple model of confinement}

In this section we investigate an explicit model,
namely a massive scalar field defined on a flat spacetime 
$R\times R^3$ in an external static field described by a   
confining potential which is asymptotically exponential in two 
dimensions. In the Euclidean version, we may compactify the ``time''
 coordinate and the $zeta$ spatial coordinate, assuming periodic 
boundary conditions with periods $\beta$ and $l$, respectively. As
 a result, the relevant operator reads
\beq
L=-\frac{d^2}{d^2\tau}-\frac{d^2}{dz^2}+H_2+M^2\,,\hs 
H_2=-\lap_2+V(r)\,,\hs V(r)=g^2e^{\alpha^2r^2}\,,
\label{l}
\eeq
$g$ and $\alpha$ being dimensional parameters. Making use of Poisson's
 re-summation formula, the heat trace can be written as 
\beq
\Tr e^{-tL}=\frac{S\,e^{-tM^2}}{4\pi t}\,\Tr e^{-t H_2}+...\,,
\eeq
where $S=\beta\, l$ and the dots stand for exponentially small terms in the parameter $t$  

In this model, the zeta function can be computed by using 
the method described in the previous section, but one now
obtains an expression which is different from  
(\ref{zetaY}), since the potential is defined only on  
$R^2$ and one needs to take the factor $e^{-tM^2/t}$ into account. 
As a result, we get 
\beq 
\zeta(s|L)&\sim&\frac{S}{(4\pi)^2\Ga(s)}\,\sum_n
\int_{0}^{\ii}\,dt\,t^{s+n-3}\int_{R^2}\,dx\,\tilde b_n(x)e^{-tV(r)}
\nn\\
&=&\sum_n\,\frac{\Ga(s+n-2)}{(4\pi)^2\Ga(s)}\,
\int_{R^2}\,dx\,\tilde b_n(x)[V(r)]^{-(s+n-2)}\,,
\label{zeZ}\eeq
where the $\tilde b_n$ are related to the  $b_n$ in (\ref{bn}) by
\beq 
\tilde b_n=\sum_{j+k=n}\,\frac{(-1)^kb_jM^{2k}}{k!}\,,\hs n\geq2\,,
\hs\hs\tilde b_0=1\,,\hs\tilde b_1=-M^2\,.
\label{}\eeq
The $\tilde b_n$ have again the same structure as in Eq.~(\ref{bn}),
but now $q$ can vanish, namely
\beq 
\tilde b_n=\sum_{pq}\,\tilde C^n_{pq}r^pa^qe^{qbr^2}\,,
  \hs\hs0\leq p\leq2(n-1)\,,\hs 0\leq q<n\,,\hs n\geq0\,,
\label{tbn}\eeq
\beq 
\tilde C^n_{00}=\frac{(-1)^nM^{2n}}{n!}\,.
\label{}\eeq
The $b_n$ coefficients which appear in Eq.~(\ref{tbn}) can be 
evaluated by making use of (\ref{G0}-\ref{G}); the first non-trivial ones read
\beq 
b_2&=&-\frac{2g \alpha e^{\alpha r^2}}{3}(1+\alpha r^2)\,,
\nn\\
b_3&=&-\frac{4g\alpha^2e^{\alpha r^2}}{15}(2+4\alpha r^2+\alpha^2 r^4)
        +\frac{g^2 \alpha^2 e^{2\alpha r^2}}{3}\,,
\nn\\
b_4&=&-\frac{8g \alpha^3 e^{\alpha r^2}}{105}
(6+18\alpha r^2+9\alpha^2 r^4+\alpha^3 r^6)
        +\frac{2g^2 \alpha^2 e^{2 \alpha r^2}}{45}(7+38\alpha r^2+21 
\alpha^2 r^4)\,,
\label{}\eeq 
from which we can read off the $C^n_{pq}$ coefficients up to $n=4$.

By integrating (\ref{zeZ}),  the  non-holomorphic 
contribution to the zeta function reads
\beq 
\ze(s|L)&=&\frac{S}{16\pi\Ga(s)}\,\sum_{n\geq0;pq}\,\tilde C^n_{pq}\,
\frac{\Ga(s+n-2)\Ga(1+p/2)\,a^{-(s+n-q-2)}}
{b^{1+p/2}\,(s+n-q-2)^{1+p/2}}\,.
\label{}\eeq
Since in our specific example $p$  is even, 
the zeta function has only poles of oder $p/2$.
In particular,  in a neighbourhood  of $s=0$ the pole structure is
\beq 
\ze(s|L)&=&\frac{S}{16\pi \alpha}\aq
\frac{M^4}{2s}
+\sum_{n=3}^6\,\frac{\tilde C^n_{2,n-2}\Ga(n-2)}{\alpha s}
+2\sum_{n=3}^6\,
\frac{\tilde C^n_{4,n-2}\Ga(n-2)}{\alpha^2 s^2}
\cq+\cdots
\label{}\eeq

Thus, the zeta function $\zeta(s|L)$ is {\it not} regular at the 
origin: a pole of second order appears. Within a physical context
(restricted most of the time to the realm of pseudo-differential 
operators in compact domains), this is a very unusual behaviour 
for the zeta function \cite{byts01}-\cite{byts02}.
In these cases, as far as the one-loop effective action is concerned, 
the otherwise well established zeta function regularisation procedure
needs to be modified \cite{byts03b,cogn04}. 

Our proposal, which extends in a natural way the one formulated
 in \cite{byts01}-\cite{byts02},  
consists in the introduction of an additional spectral function 
which depends on the order of the pole at the origin of the initial
zeta function. Thus, in the case of a pole of order $N$, 
the auxiliary spectral function reads 
\beq
\omega(s)=s^N \zeta(s|L)\,,
\eeq
and the definition of the regularised determinant is generalised as
\beq
\ln \det \frac{L}{\mu^2}=-\frac{1}{(N+1)!}\lim_{s \rightarrow 0}
\frac{d^{N+1}}{ds^{N+1}}\aq \mu^{2s} \omega(s) \cq\,,
\label{genZ}
\eeq
with the normalisation chosen in such a way that when $\zeta(s|L)$
is regular at the origin, one does recover the ordinary definition of
regularised functional determinant. This is an essential condition 
in order to preserve the well established properties defining the
zeta function regularisation procedure.

Recalling our example before, we have seen that in this model a 
second-order pole will generically appear. Then, the new spectral 
function, which is regular at the origin, will be given by
\beq
\omega(s)=s^2 \zeta(s|L)\,.
\eeq
We correspondingly define
\beq
\ln\det\frac{L}{\mu^2}=-\frac{1}{3!}\lim_{s\to0}
\frac{d^3}{ds^3}\aq\mu^{2s}\omega(s)\cq\,.
\eeq
It has to be mentioned here that, within the context of a general 
theory of regularised products (see e.g. \cite{Illies}), in the case 
when the related zeta 
function is not holomorphic at the origin but has a first-order pole
---and we have stressed this to happen
when logarithmic terms are present in the heat-trace asymptotics--- 
a new definition of regularised product was proposed  recently
\cite{hirano}. It reads
\beq
\prod_{k=1}^{\ii}\la_k\equiv\exp{\aq - 
\Res(\frac{\zeta(s)}{s^2})_{s=0}\cq}\,,
\hs\hs\ze(s)=\sum_{k=1}^\ii \lambda_k^{-s}\,.
\label{H}   
\eeq
Recalling the definition of residue, it is straightforward to conclude 
that this prescription is equivalent to ours, Eq.~(\ref{genZ}). This
is a further consistency check and inscribes our result in a
very general context.

We conclude with the following remark. The one-loop renormalisation 
group equations associated with the presence of the renormalisation 
scale $\mu$ can be treated along the same lines as in 
Ref.~\cite{cogn04}. This shows both the power and flexibility of 
the zeta-function method to easily cope with non-standard
and unexpected situations, without ever losing contact with the 
fundamental issue of its applicability to actual physical problems.
This means, in particular, that the results obtained with the method 
must be checked to be physically meaningful and to reproduce measured 
experimental values. 

\section{Conclusion}
In this paper we have considered several examples of the determination
 of heat-kernel traces associated with operators of  Laplace type
defined on non-compact domains.
For the sake of simplicity, we have restricted our analysis to R$^d$
and to analytic but confining potentials, thus dealing with
discrete spectra only. However, the adequacy of the procedure to treat more
general settings has been exhibited. In particular, although for the 
sake of simplicity we have postponed the treatment of the case
when $d/Q$ is non-integer, with some extra effort this can be dealt 
with along the same lines. New branching  points appear there.

We have shown that for confining potentials of exponential
behaviour at infinity, the asymptotics of the heat-kernel trace
contain generically logarithmic terms. As a consequence,
the meromorphic structure of the associated
zeta function develops higher-order poles as well as branching points. 
In particular, we have exhibited some cases where the zeta
function is not regular at the origin.

In these situations, one is confronted with the non-trivial task of
having to define the corresponding regularised functional determinant 
or the one-loop effective potential. In fact, in the example of an
apparently reasonable model of confinement, constructed by means of
an asymptotically exponential potential, we
have proven that the meromorphic continuation of the zeta function
already develops a higher-order pole at the origin. This, as far
as we know, is an absolutely novel finding in the field,
even more since it comes from such an apparently harmless model.

In order to deal with these special cases, we have proposed a
generalisation of the zeta function regularisation procedure,
consisting in the introduction of a new, auxiliary
zeta function which is still regular at the origin,
together with a correspondingly new definition for the
zeta-regularised determinant (thus extending Ray and Singer's
definition). This general prescription ---which naturally extends
particular cases already considered by some of us before 
\cite{byts01}-\cite{byts02}--- turns out to be equivalent to the one 
recently proposed by Hirano et al.~\cite{hirano} in a more generic
context of a theory of regularised products. In accordance
with our fundamental aim never to abandon the already established 
connections with the physical world (e.g., the many
uses of zeta regularisation in experimental physics), all the
new quantities have been defined in such a way  as to recover the  
celebrated results of zeta-function regularisation in the 
absence of  poles at the origin.

Still pending is the task to construct an explicit general theory
to deal with the whole class of Laplacian like operators in
non-compact domains, what we leave for further work.

\ack

We thank the referee for valuable remarks and criticisms which 
have resulted in a significant improvement of this paper.
This paper is an outcome of the collaboration program INFN 
(Italy)--DGICYT (Spain). EE has been also supported by DGICYT 
(Spain), project BFM2003-00620, and  AGAUR (Generalitat de 
Catalunya), contract 2005SGR-00790.

\appendix

\section{Appendix: Some useful formulae}

Here we list some expressions that have been employed in the text. We
start with the incomplete gamma function, useful in order to reveal
the presence of logarithmic terms in the heat-kernel trace expansion.
Its definition reads 
\beq 
\Gamma(s,t)=\int_t^\ii dy\, y^{s-1}e^{-y}
=\Ga(s)-\frac{t^s}{s}-t^s\,\sum_{r=1}^\ii\,\frac{t^n}{n!(s+n)},
\label{a5} \eeq 
and thus 
\beq
\Gamma(0,t)=-\ln t-\gamma-t-\frac{t^2}{4}+O(t^3)\,, 
\label{a6}\eeq
$\gamma$ being Euler's constant. 
Taking the derivative of order $n$ of 
$\Gamma(s,t)$ with respect to $s$, one gets
\beq
\frac{d^n}{ds^n}\Gamma(s,t)=\int_t^\ii dy\,\,(\ln y)^n y^{s-1}e^{-y}\,,
\label{a7} \eeq 
from which it follows that
\beq 
\int_1^\ii\,dy\,(\ln y)^ny^{s-1} e^{-ty}
&=&\frac{d^n}{ds^n}\,\frac{\Ga(s,t)}{t^s}
\nn\\
&=&\sum_{k=0}^n\,f_k(s)\Ga(s)\,\frac{\at\ln t\ct^k}{t^s}
-\frac{(-1)^nn!}{s^{n+1}}+O(t),
\label{XXX}\eeq
where the $f_k(s)$ are computable functions. In particular, 
\beq 
f_n(s)=(-1)^n\,,\hs\hs f_{n-1}(s)=(-1)^{n+1}\psi(s)\,,
\label{}\eeq
$\psi(s)$ being the digamma function.
In the limit $s\to0$ we finally have
\beq 
\int_1^\ii\,dy\,(\ln y)^ny^{-1} e^{-ty}
=\sum_{k=0}^{n+1}\,c_k\at\ln t\ct^k+O(t)\,,
\label{bbb2}\eeq
where
\beq 
c_{n+1}=\frac{(-1)^{n+1}}{n+1}\,,\hs\hs c_n=(-1)^{n+1}\ga\,.
\label{}\eeq

With a view to the analytic continuation of zeta functions, the
following formulas are useful too.
If $\alpha$ is a complex number with positive real part, and $\beta$
such that Re $\beta  >-1$, one has 
\beq \int_0^1 dt \, t^{\alpha-1}(\ln t)^\beta
= \frac{(-1)^\beta \Gamma(\beta+1)}{\alpha^{\beta+1}}\,.
\label{a1} \eeq 
To prove this, it is sufficient to perform the change
of variable $u=-\ln t$ and recall the definition of the Euler gamma
function.

Furthermore, for the asymptotics of the spectral density for large
$\lambda$, one has
\beq
t^{-s}=\int_0^\ii  e^{-t\lambda} \frac{\lambda^{s-1}}{\Gamma(s)}\,.
\label{a2}
\eeq
Taking the derivative with respect to $s$, 
\beq
\ln t \, t^{-s}=\int_0^\ii  e^{-t\lambda} \lambda^{s-1}
\aq -\frac{\ln \lambda}{\Gamma(s)}+\frac{\Gamma'(s)}{\Gamma(s)}\cq \,,
\label{a3}
\eeq
and
\beq
\ln^2 t \, t^{-s}=\int_0^\ii e^{-t\lambda} \lambda^{s-1}
\aq \frac{\ln^2 \lambda}{\Gamma(s)}+2\frac{\Gamma'(s)}{\Gamma^2(s)} \ln
\lambda  -  \frac{\Gamma''(s)}{\Gamma(s)}-\frac{\Gamma'(s)}{\Gamma^2(s)} \cq
\,.
\label{a4}
\eeq
The above identities are compatible with the Karamata Tauberian 
theorems, which can be stated as follows. Suppose we deal with
\beq
\int_0^\ii  e^{-t\lambda} d N(\lambda)=K(t)\,.
\label{a9}
\eeq
i) If
\beq
K(t)\simeq A \, t^{-r}\,, \hs t \rightarrow 0\,,
\eeq
then
\beq
N(\lambda)\simeq A \frac{\lambda^{r}}{\Gamma(r+1)}\,,
\hs \lambda \rightarrow \ii\,.
\eeq
ii) If
\beq
K(t)\simeq A \, t^{-r}\ln^N t\,, \hs t \rightarrow 0\,,
\eeq
then
\beq
N(\lambda)\simeq A \frac{\lambda^r \ln^N \lambda}{\Gamma(r+1)}\,,
\hs \lambda \rightarrow \ii\,.
\eeq


\begin{thebibliography}{10}

\bibitem{eliz94b}
E.~Elizalde, S.D.~Odintsov, A.~Romeo, A.A.~Bytsenko and S.~Zerbini,
{\em Zeta Regularisation Techniques with Applications},
World Scientific, Singapore (1994).

\bibitem{eli10}
E.~Elizalde, {\em Ten Physical Applications of Spectral Zeta Functions},
Springer Verlag, Berlin (1995).

\bibitem{byts96-266-1}
A.A.~Bytsenko, G.~Cognola, L.~Vanzo and S.~Zerbini,
 Phys.~Rept. {\bf 266}, 1 (1996).

\bibitem{bek4}
M. Bordag, E. Elizalde and K. Kirsten, J. Math.
Phys. {\bf 37}, 895 (1996); M. Bordag, E. Elizalde, K. Kirsten and
S. Leseduarte,  Phys. Rev. {\bf D56}, 4896 (1997); E. Elizalde, M.
Bordag and K. Kirsten, J. Phys. {\bf A31}, 1743 (1998); M. Bordag,
E. Elizalde, B. Geyer and K. Kirsten, Commun. Math. Phys. {\bf 179},
215 (1996).
\bibitem{byts03b}
A.A.~Bytsenko, G.~Cognola, E.~Elizalde, V.~Moretti and S.~Zerbini,
{\em Analytic  Aspects of Quantum Fields}, World Scientific, Singapore (2003).

\bibitem{z1}
D.B.~Ray and I.M.~Singer, Ann. Math. {\bf 98}, 154 (1973).

\bibitem{z2}
S.W. Hawking, Commun. Math. Phys. {\bf 55}, 133 (1977).

\bibitem{dowk76-13-3224}
J.S.~Dowker and R.~Critchley, Phys.~Rev. {\bf D13}, 3224 (1976).

\bibitem{dewi65b}
B.S.~DeWitt, {\em The Dynamical Theory of Groups and Fields},
Gordon and Breach, New York (1965).

\bibitem{seel67-10-172}
R.T.~Seeley,
 Am.~Math.~Soc.~Prog.~Pure Math. {\bf 10}, 172 (1967).

\bibitem{kirsten00}
K. Kirsten, {\em Spectral Functions in Mathematics and Physics},
Chapman and Hall, CRC, London (2000).

\bibitem{vass03}
D.V.~Vassilevich,  Phys. Rept. {\bf 388}, 279 (2003).


\bibitem{parker1}
L. Parker and D.J. Toms, Phys. Rev. {\bf D31}, 953 (1985).


\bibitem{parker2}
J. Jack and L. Parker, Phys. Rev. {\bf D31}, 2439 (1985).

\bibitem{bord96}
M. Bordag, K. Kirsten and J.S. Dowker,  Commun. Math. Phys. {\bf
182}, 371 (1996).


\bibitem{cogn97}
G. Cognola and S. Zerbini, Lett. Math. Phys. {\bf 42}, 95 (1997).


\bibitem{byts97}
A.A. Bytsenko, G. Cognola and S. Zerbini, 
 J. Phys.  {\bf A30}, 3543 (1997);
A.A. Bytsenko, G. Cognola and S. Zerbini,
Class. Quantum Grav. {\bf 14}, 615 (1997).


\bibitem{grubb}
G. Grubb,  Nucl. Phys. Proc. Suppl. {\bf B104}, 71 (2002).

\bibitem{byts01}
A.A. Bytsenko, E. Elizalde and S. Zerbini,
Phys. Rev. {\bf D64}, 105024 (2001).

\bibitem{eejpa01}
E. Elizalde,  J. Phys. {\bf A34}, 3025 (2001).

\bibitem{byts02}
A.A. Bytsenko, A.E. Gon\c calves and S. Zerbini,
 Mod. Phys. Lett. {\bf A16}, 1479 (2001).


\bibitem{nash}
C. Nash, Ad. Applied Maths. {\bf 6}, 436 (1985).


\bibitem{voros}
A. Voros, J. Phys. {\bf A33}, 7423 (2000).


\bibitem{dowker}
J.S. Dowker and K. Kirsten, 
Analysis Appl. {\bf 3}, 45 (2005).


\bibitem{callias}
C. Callias,  Commun. Math. Phys.  {\bf 88}, 357 (1983); C.
Callias,  Commun. Partial Diff. Eqs.  {\bf 13}, 1113 (1988).



\bibitem{kirsten}
K. Kirsten, P. Loya and J. Park,
J. Math. Phys. {\bf 47} 043506 (2006).


\bibitem{Illies}
G. Illies,  Commun. Math. Phys. {\bf 220}, 69 (2001).


\bibitem{jorgenson}
J. Jorgenson, S. Lang and D. Goldfeld, {\em Explicit  formulas},
 LNM {\bf 1593}, Berlin, Springer (1994);
J. Jorgenson and S. Lang, {\em Basic Analysis of regularized series and products},
 LNM {\bf 1564}, Berlin, Springer (1994).

\bibitem{manin}
Yu. Manin, Asterisque {\bf 228}, 121 (1995).


\bibitem{dencker}
N. Dencker,  Ark. Math.  {\bf 24}, 59 (1986).


\bibitem{simon}
B. Simon,  Ann. Phys. (NY) {\bf 146}, 209 (1983).



\bibitem{muller}
W. M\"{u}ller, Commun. Math. Phys. {\bf 192}, 309 (1998).


\bibitem{solo}
S. Solodhukin, Nucl. Phys.  {\bf B541}, 401 (1999).


\bibitem{irina}
M. Bordag and D.V. Vassilevich,  J. Phys. {\bf A32}, 8247 (1999);
M. Bordag, I.G. Pirozhenko and V.V. Nesterenko,
J. Phys. {\bf A38}, 11027 (2005).


\bibitem{nail}
N.R. Khusnutdinov,
 Phys. Rev. {\bf D73} 025003 (2006).


\bibitem{cramer}
H. Cramer,  Math. Z. {\bf 4}, 104 (1919);
J. Jorgenson and S. Lang,  Math. Ann.  {\bf 297}, 383 (1993).


\bibitem{rosu}
H.C. Rosu, Mod. Phys. Lett. {\bf A18}, 1205 (2003).

\bibitem{sierra}
G. Sierra, 
J .Stat. Mech. {\bf 0512}, P12006 (2005).


\bibitem{cogn04}
G. Cognola and S. Zerbini, Phys. Rev. {\bf D69}, 024004 (2004).

\bibitem{hirano}
M. Hirano, N. Kurokawa, M. Wakayama,
 J. Ramanujan Math. Soc. {\bf 18}, 195 (2003).

\end{thebibliography}
\end{document}